\documentclass[11pt,a4paper]{article}
\usepackage{jheppub}
\usepackage{amsfonts}
\usepackage{amssymb}
\usepackage{slashed}
\usepackage{CJK}
\usepackage{epsfig}

\def\Spaa #1{\left\langle #1 \right \rangle}
\def\Spbb#1{\left[ #1 \right]}
\def\Spab#1{\left\langle #1 \right]}
\def\Spba#1{\left[ #1 \right \rangle}
\def\bket#1{\left| #1\right]}
\def\ket#1{\left| #1\right\rangle}

\newcommand{\bea}{\begin{eqnarray}}
\newcommand{\eea}{\end{eqnarray}}
\newcommand{\la}{\begin{equation}}
\newcommand{\ra}{\end{equation}}
\newcommand{\bean}{\begin{eqnarray*}}
\newcommand{\eean}{\end{eqnarray*}}

\def\W #1{\widetilde{#1}}
\def\WH #1{\widehat{#1}}

\def\Label#1{\label{#1}%
  \smash{\hbox to0pt{\raise1ex\hbox{\tiny[#1]}\hss}}}

\allowdisplaybreaks
\title{Boundary Contributions Using Fermion Pair Deformation}

\author[a,b,c]{Bo Feng}
\author[c,d]{Zhibai Zhang}
\affiliation{$^a$\small Zhejiang Institute of Modern Physics, Zhejiang
University, Hangzhou, 310027, P. R. China\\$^b$\small Center of
Mathematical Science, Zhejiang University, Hangzhou, China \\
$^c$\small Kavli Institute for Theoretical Physics China, CAS,
Beijing 100190, China \\
$^d$\small The Graduate School and University Center, The City University of New York 365 Fifth Avenue, New York NY 10016, USA}
\emailAdd{b.feng@cms.zju.edu.cn}
\emailAdd{zzhang2@gc.cuny.edu}

\abstract{Continuing the study of boundary BCFW recursion relation
of tree level amplitudes initiated in \cite{Feng:2009ei}, we
consider  boundary contributions coming from fermion pair
deformation. We present the general strategy for these boundary
contributions and demonstrate calculations using two examples, i.e,
the standard QCD and deformed QCD with anomalous  magnetic momentum
coupling. As a by-product, we have extended BCFW recursion relation
to off-shell gluon current, where because off-shell gluon current is
not gauge invariant, a new feature must be cooperated.

} \keywords{Amplitude Calculation}
\begin{document}

\maketitle


\section{Motivations}
The calculation of amplitudes is always a key problem in  quantum
field theories. The familiar method of Feynman diagrams faces a lot
of challenges when the process involves a lot of external particles
or couples to gauge theory, thus  more efficient new methods are
wanted.

There are many novel approaches\footnote{For some  reviews, see
\cite{mangparke,Dixon:1996wi, Peskin}.} for calculating amplitudes
efficiently in the past two decades, such as  the spinor method, the
color ordering technique, the twistor method initiated in
\cite{Witten:2003nn, Hodges:2005bf, ArkaniHamed:2009si}, the CSW
method \cite{Cachazo:2004kj} using the compact MHV amplitudes
\cite{Parke} as vertexes, the Grassmannian method
\cite{ArkaniHamed:2009dn} and the Wilson Loop method
\cite{Alday:2007hr}. Along these breakthroughs, a new on-shell
recursion relation for tree level amplitude  was found in
\cite{Britto:2004ap} and  proven in \cite{ Britto:2005fq} shortly.
The on-shell recursion relation can be schematically written  as
\bea A_n=\sum_{L}\sum_{helicity} A_L\cdot\frac{1}{P^2}\cdot
A_R~~~\Label{1.1}\eea
where $A_n$ is the tree amplitude involving $n$ gluons, $A_L$ and
$A_R$ are on-shell sub amplitudes and $\frac{1}{P^2}$ is
corresponding pole. Although the original recursion relation is for
gauge theory, very rapidly it was understood  that the validity of
BCFW recursion relation relies on some general complex analytic
structures of tree-level amplitudes. Thus it is extended to other
field theories, including some effective theories, based on the same
analysis\footnote{A recent review can be found in
\cite{Brandhuber:2011ke}.}.

With these generalizations, the important role of the large $z$
behavior\footnote{A very nice analysis of large $z$ behavior can be
found in \cite{ArkaniHamed:2008yf, Cheung:2008dn}.} of amplitude
under the deformation $p_i\rightarrow p_i-zq,\;p_j\rightarrow
p_j+zq$ with $q^2=p_i\cdot q=p_j\cdot q=0$ has been realized. The
reason is that we need to use the contour integration $\oint
{dz\over z} A(z)$ to derive the recursion relation, where $A(z)$ is
the rational function of $z$ obtained from original amplitude with
deformation. However, if under the limit $z\to \infty$, $A(z)\to
C_0+ C_1 z+...C_k z^k$ with $C_0\neq 0$, the contour $\oint {dz\over
z} A(z)\neq 0$, i.e., it has nonzero boundary  contributions at
infinity. Unlike the pole at finite $z$, where residue can be
inferred from factorization property, we do not know how to describe
 boundary contributions from the first principle, thus in many practices
  we ask the vanishing behavior $A(z\to \infty)\to 0$ to avoid the
trouble.

Although the vanishing condition makes the derivation of recursion
relation simpler, it constraints the scope of application of
recursion relation, such as $\phi^4$ theory and theories with Yukawa
coupling. Thus it is very interesting to generalize the on-shell
recursion relation to cases where there are nonzero boundary
contributions. Some progresses along this direction have been given
in \cite{Feng:2009ei, Feng:2010ku, Benincasa:2011kn} where two
methods have been proposed to investigate boundary contributions.
The first method is to analyze Feynman diagrams so we can isolate
boundary contributions. For many theories, only small part of
Feynman diagrams gives contributions and their direct calculations
are not so difficult. The second method is to translate information
of boundary contributions to the information of zero of amplitudes,
i.e., the number  of zero and their explicit values. Comparing these
two methods, the second one is general, but difficult to calculate
while the first one is more intuitive.

In this paper, we will continue our study of the boundary BCFW
recursion relation
\bea A_n=\sum A_L\cdot \frac{1}{P^2}\cdot A_R+A_b\eea
where $A_b$ is the boundary contribution part. The complexity of
boundary contributions increases with the complexity of wave
functions of deformed external particles. While wave function of
scalar particles is simple, the wave function of fermions and gluons
are not. We will focus on the fermion deformation in this paper, but
our method could be generalized to gluons and gravitons.

This paper is organized as follows. To prepare calculations in
section three and four,  we discuss the off-shell gluon current in
section two. After reviewing the Berends-Giele off-shell recursion
relation \cite{Berends:1987me}, we present a new recursion relation
using the BCFW-deformation. Because the off-shell current is not
gauge invariant, the new recursion relation need to sum up four
helicity states instead of just two physical helicity states met in
usual on-shell recursion relation.  In section three, using Feynman
diagrams we isolated  boundary contributions in QCD with deformed
fermion pair. Having this experience, in section four we studied the
modified QCD theory with anomalous magnetic momentum coupling
presented in \cite{Larkoski:2010am} and write down the corresponding
boundary BCFW recursion relation for a special helicity
configuration. Finally, a brief summary is given in section five.

\section{Calculations of off-shell gluon currents}

In this section, we will revisit the calculation of color-ordered
off-shell current $J^{\mu}(1,2,...,k)$ of gauge theory, which  will
be useful when we discuss  possible boundary contributions in BCFW
on-shell recursion relation for theories coupled with gauge theory.
Different from on-shell amplitude, the off-shell current
$J^{\mu}(1,2,...,k)$ is gauge dependent as there is a leg
un-contracted with physical polarization vector. The gauge freedom
comes from several places. The first gauge freedom is the choice of
a null reference momentum when we define the physical polarization
vector for an external on-shell gluon
\bea \epsilon_{i\mu}^{+}= \frac{\Spab{ r_i|\gamma_{\mu}|p_i}}{\sqrt2
\Spaa{r_i|p_i}},~~~~~\epsilon_{i\mu}^{-}= -\frac{\Spba{
r_i|\gamma_{\mu}|p_i}}{\sqrt2 \Spbb{r_i|p_i}}~~~~~\label{2.1}\eea
where the $p_i$ is the momentum of the $i$-th gluon and $r_i$ is the
null reference momentum.  The second gauge freedom is the choice of
gluon propagator
\bea
D^{\mu\nu}(p)=\frac{-i}{p^2}\left(g^{\mu\nu}-(1-\xi)\frac{p^{\mu}p^{\nu}}{p^2}\right)~~~~\Label{2.2}\eea
where $\xi=1$ is the familiar Feynman gauge.

Besides the physical polarization vector defined in (\ref{2.1}),
there are other two polarization vectors we can define
\bea \epsilon_\mu^L =
p_i,~~~~\epsilon_\mu^T={\Spab{r_i|\gamma_\mu|r_i}\over 2 p_i\cdot
r_i}~~~\Label{2.1.1}\eea
Using the {\bf Fierz rearrangement}
\bea \Spba{i|\gamma^\mu|j}\Spba{k|\gamma_\mu|l}=2
\Spbb{i|k}\Spaa{l|j}~~~\Label{Fierz}\eea
we find that
\bea 0 & = & \epsilon^+\cdot \epsilon^+= \epsilon^+\cdot
\epsilon^L=\epsilon^+\cdot \epsilon^T=\epsilon^-\cdot
\epsilon^-=\epsilon^-\cdot \epsilon^L=\epsilon^-\cdot \epsilon^T
=\epsilon^T\cdot \epsilon^T=\epsilon^L\cdot \epsilon^L\nonumber \\
1 &= & \epsilon^+\cdot \epsilon^- =\epsilon^L\cdot
\epsilon^T~~~\Label{2.1.2}\eea
Thus these four vectors give a basis in the four-dimension space
time and we have
\bea g_{\mu\nu}= \epsilon_\mu^+ \epsilon_\nu^- +\epsilon_\mu^-
\epsilon_\nu^+ +\epsilon_\mu^L \epsilon_\nu^T+\epsilon_\mu^T
\epsilon_\nu^L ~~~\Label{2.1.3}\eea
Formula (\ref{2.1.3})  will be important for our late calculation.

The off-shell current can be calculated using Feynman diagrams, but
there is a better way to calculate using the Berends-Giele off-shell
recursion relation \cite{Berends:1987me}. To do so, we need
following color-ordered three-leg vertex $V_3$ and four-leg vertex
$V_4$ given as
 \bea
V_{3}^{\mu\nu\rho}\left(p,q\right) & = & \frac{i}{\sqrt{2}}\left(\eta^{\nu\rho}\left(p-q\right)^{\mu}+2\eta^{\rho\mu}q^{\nu}-2\eta^{\mu\nu}p^{\rho}\right)\nonumber\\
V_{4}^{\mu\nu\rho\sigma} & = &
\frac{i}{2}\left(2\eta^{\mu\rho}\eta^{\nu\rho}-\eta^{\mu\nu}\eta^{\rho\sigma}-
\eta^{\mu\sigma}\eta^{\nu\rho}\right)~~~~\Label{2.3}\eea
Using above definition, the color-ordered off-shell recursion
relation is given by\footnote{The factor $\frac{-i}{P_{1,k}^{2}}$
tells us that the formula uses the Feynman propagator.}
 \bea
J^{\mu}\left(1,2,...,k\right) & = & \frac{-i}{P_{1,k}^{2}}\left[\sum_{i=1}^{k-1}V_{3}^{\mu\nu\rho}\left(P_{1,i},P_{i+1,k}\right)J_{\nu}\left(1,...,i\right)J_{\rho}\left(i+1,...,k\right)\right.\nonumber\\
 &  & \left.+\sum_{j=i+1}^{k-1}\sum_{i=1}^{k-2}V_{4}^{\mu\nu\rho\sigma}
J_{\nu}\left(1,...,i\right)J_{\rho}\left(i+1,...,j\right)J_{\sigma}
\left(j+1,...,k\right)\right]\Label{2.4}\eea
where $P_{i,j}=p_{i}+p_{i+1}+\cdots+p_{j}$ and $P_{1,k}$ is the
momentum carried by the off-shell leg. A graphic description of the
off-shell recursion relation is showed in Fig \ref{Fig:division-m}.
As a recursion relation, (\ref{2.4}) has a starting point
$J^{\mu}(1)=\epsilon^{\pm\mu}(p_1)$, which is the current with only
one on-shell gluon.

   \begin{figure}[hbt]
  \centering
  \scalebox{1.13}[1.13]{\includegraphics{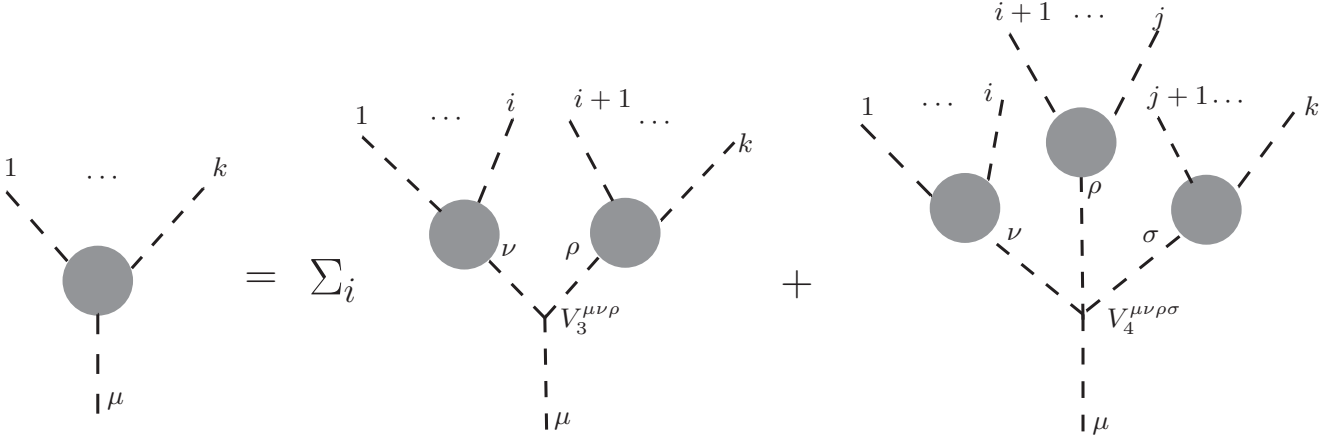}}\quad
  \caption{A graphic description for the off-shell recursion relation of  gluon current \Label{Fig:division-m}}
\end{figure}

Although off-shell recursion relation is a better organization than
 Feynman diagrams, the calculation of
$J^{\mu}\left(1,2,...,k\right)$ with general helicity configuration
is still very complicated and the result is highly non-compact and
gauge dependent. Nevertheless, the gauge freedom indicates that
there are two helicity configurations of which  results are compact
under proper gauge choices.  The first case is that all helicities
in the current are the same, for example  with positive helicities,
and the result is given by
\bea J^{\mu}\left(1^{+},2^{+},...,k^{+}\right)=\frac{\Spaa{r|\gamma^{\mu}\slashed{P}_{1,k}|r}}{\sqrt{2}\Spaa{r1}\Spaa{12}\cdots\Spaa{k-1,k}\Spaa{kr}}\Label{2.6}
\eea
where all  reference momenta of gluons  are chosen to be $r$. The
second case is that only the first gluon has negative helicity, and
the current is given by
\bea
J^{\mu}\left(1^{-},2^{+},3^{+},...,k^{+}\right)=\frac{\langle1|\gamma^{\mu}\slashed{P}_{2,k}|1\rangle}{\sqrt{2}\langle12\rangle\cdots\langle
k1\rangle}\sum_{i=3}^{k}\frac{\langle1|\slashed{k}_{i}\slashed{P}_{1,i}|1\rangle}
{P_{1,i-1}^{2}P_{1,i}^{2}}\Label{2.7}\eea
where  reference momenta are chosen as following: $r_{1}=p_{2},\:
r_{2}=r_{3}=\cdots=r_{k}=p_{1}$. It is important to notice that for
a relatively simple result,  gauge choices must be made as above.

To illuminate above discussions, we give the derivation of 4-point
current $J^{\mu}\left(1^{-},2^{+},3^{+},4^{+}\right)$. To simplify
the writing, we define functions $I^{\mu}[\cdot,\cdot]$ and
$I^{\mu}[\cdot,\cdot,\cdot]$ as following:
\bea
I^{\mu}\left[J(1,2,...,i),J(i+1,...,k)\right]=\frac{-i}{P_{1,k}^2}V_3^{\mu\nu\rho}(P_{1,i},P_{i+1,k})J_{\nu}(1,2,...,i)J_{\rho}(i+1,...,k)\nonumber\\
I^{\mu}\left[J(1,2,...,i),J(i+1,...,j),J(j+1,...,k)\right]=\frac{-i}{P_{1,k}^2}V_{4}^{\mu\nu\rho\sigma}
J_{\nu}\left(1,...,i\right)J_{\rho}\left(i+1,...,j\right)J_{\sigma}
\left(j+1,...,k\right)~~\Label{2.8}\eea
Thus the off-shell recursion relation given in (\ref{2.4}) could be
written as
\bea J^{\mu}\left(1,2,...,k\right)=
\sum_{i}I^{\mu}[J(1,2,...,i),J(i+1,...,k)]+\sum_{i,j}I^{\mu}[J(1,2,...,i),J(i+1,...,j),J(j+1,...,k)]~~\Label{2.9}\eea
With this notation, the 4-point current $J^{\mu}(1,2,3,4)$ could be
written recursively as
\bea
J^{\mu}\left(1,2,3,4\right) & = & I^{\mu}\left[J(1),J\left(2,3,4\right)\right]
+I^{\mu}\left[J\left(1,2\right),J\left(3,4\right)\right]+I^{\mu}\left[J\left(1,2,3\right),J(4)\right]\nonumber\\
 &  & +I^{\mu}\left[J(1),J(2),J\left(3,4\right)\right]+I^{\mu}
\left[J(1),J\left(2,3\right),J(4)\right]+I^{\mu}\left[J\left(1,2\right),J(3),J(4)\right]\eea
For the helicity configuration
$\left(1^{-},2^{+},3^{+},4^{+}\right)$ we choose the reference
momenta as $r_{1}=p_{2},\: r_{2}=r_{3}=r_{4}=p_{1}$, then it is not
difficult to check that following four terms vanish
\bea
&&I^{\mu}\left[J\left(1^{-},2^{+}\right),J\left(3^{+},4^{+}\right)\right]
=I^{\mu}\left[J(1^{-}),J(2^{+}),J\left(3^{+},4^{+}\right)\right]
\nonumber\\&&=
I^{\mu}\left[J(1^{-}),J\left(2^{+},3^{+}\right),J(4^{+})\right]
=I^{\mu}\left[J\left(1^{-},2^{+}\right),J(3^{+}),J(4^{+})\right]=0\eea
while the other two non vanishing terms are given as
\bea
I^{\mu}\left[J(1^{-}),J\left(2^{+},3^{+},4^{+}\right)\right]
& = & -\frac{\langle1|3+4|2]\langle1|\gamma^{\mu}\slashed{k}_{234}|1\rangle}{\sqrt{2}s_{1234}
s_{12}\langle23\rangle\langle34\rangle\langle41\rangle}\\
I^{\mu}\left[J\left(1^{-},2^{+},3^{+}\right),J(4^{+})\right] & = &
\frac{[23]\langle1|2+3|4]}{\sqrt{2}s_{12}s_{123}s_{1234}\langle14\rangle\langle23\rangle}
\langle1|\gamma^{\mu}\slashed{k}_{234}|1\rangle\eea
Adding them up we obtain
\bea
J^{\mu}\left(1^{-},2^{+},3^{+},4^{+}\right) & = & I^{\mu}\left[1^{-},\left(2^{+},3^{+},4^{+}\right)\right]+I^{\mu}\left[\left(1^{-},2^{+},3^{+}\right),4^{+}\right]\nonumber\\
 & = & \frac{\langle1|\gamma^{\mu}\slashed{k}_{234}|1\rangle}{\sqrt{2}\langle12
 \rangle\langle23\rangle\langle34\rangle\langle41\rangle}
 \left(\frac{\langle1|\slashed{3}\cdot\left(\slashed{1}+\slashed{2}+\slashed{3}\right)
 |1\rangle}{s_{12}s_{123}}+\frac{\langle1|\slashed{4}\cdot\left(\slashed{1}
 +\slashed{2}+\slashed{3}+\slashed{4}\right)|1\rangle}{s_{123}s_{1234}}\right)\eea
 which is the one given by (\ref{2.7}).

Having shown the calculation of current by off-shell recursion
relation, it is natural to ask if we can do it using the new
discovered on-shell recursion technique. In following two
subsections, we will discuss this issue.

\subsection{Recursion relation by two on-shell gluon deformation}\Label{Onshell}

The off-shell current $J^\mu(1,2,...,k)$ has $k$ on-shell gluons,
thus it is obvious that  we can take a pair of on-shell gluons to do
the BCFW-deformation and write down the corresponding BCFW recursion
relation for the current. The boundary behavior under the
deformation $\Spba{i|j}$ (i.e., the deformation $\bket{i}\to
\bket{i}-z\bket{j}$, $\ket{j}\to \ket{j}+z\ket{i}$) will be ${1\over
z}$ for the helicity configurations $(-,+),(+,+),(-,-)$ and $z^3$
for the helicity configuration $(+,-)$ \footnote{The boundary
behavior is, in fact, more subtle. For example, if $(i,j)$ are not
nearby, we will have ${1\over z^2}$ behavior for
$(-,+),(+,+),(-,-)$. But for our purpose, naive counting is enough.}
and the off-shell leg will not cause any trouble.

With above explanation, if the helicity of $(1,k)$ is
$\left(-,+\right)$, $\left(-,-\right)$ and $\left(+,+\right)$,  we
can take the deformation on $1$ and $k$
\bea |1] & \rightarrow &
|1]-z|k],\; |k\rangle  \rightarrow |k\rangle+z|1\rangle\eea
and the corresponding recursion relation is given by
 \bea
J^{\mu}\left(1,2,...,k\right) & = & \sum_{i=2}^{k-1}\sum_{h,\W h}
\left[A\left(\WH{1},...,i,\WH{P}^{h}\right)\cdot\frac{1}{P_{1,i}^{2}}\cdot J^{\mu}\left(-\WH{P}^{\W h},i+1,...,\WH{k}\right)\right.\nonumber\\
+J^{\mu}\left(\WH{1},...,i,\WH{P}^{h}\right)& &\left. \cdot\frac{1}{P_{i+1,k}^{2}}\cdot A\left(-\WH{P}^{\W h},i+1,...,\WH{k}\right)\right],~~
 (h,\W h)=(+,-),(-,+),(L,T),(T,L)~~\Label{2.16}\eea
where the graphic description is given in Fig \ref{Fig:division-22}.

There are several things we need to emphasize for the formula
(\ref{2.16}). First, since the current
$J^{\mu}\left(1,2,...,k\right)$ itself is gauge dependent, all
reference momenta in the sub-currents at the right hand side of
(\ref{2.16}) must  be the same with these at the left hand side of
(\ref{2.16}). A consequence of this requirement is that we can not
naively use results (\ref{2.6}) and (\ref{2.7}), which are results
with special choices of gauge.

Secondly, for the on-shell momentum $\WH P$ at the right hand side
of (\ref{2.16}), we must sum over four polarization vectors defined
in (\ref{2.1}) and (\ref{2.1.1}) (not just the vectors in
(\ref{2.1})) by the formula (\ref{2.1.3}) for Feynman propagator.
The reason that we can neglect the sum over vectors in (\ref{2.1.1})
for on-shell amplitude is because $\epsilon^L=\WH P$ and by the Ward
Identity, when all other particles are on-shell and with physical
polarizations, $\WH P\cdot A=0$.  Thus for  other two configurations
$(h,\W h)=(L,T),(T,L)$ in (\ref{2.16}), we have either the $\WH
P\cdot A_L=0$ or $\WH P\cdot A_R=0$, so we are left with only two
familiar  helicity configurations in BCFW recursion relation for
on-shell amplitudes. For off-shell current we are interesting in, we
do not have $\epsilon^{L,T}\cdot J\neq 0$, thus we can not neglect
the sum over $(h,\W h)=(L,T),(T,L)$. However, we will show that
usually these two terms can vanish by special choice of gauge. Also
in the practice we should use the Ward Identity to simplify the
calculation. For example,  with $(h,\tilde{h})=(T,L)$ configuration
the second term of (\ref{2.16}) vanishes according to Ward Identity
\bea A\left(-\WH{P}^{L},i+1,...,\WH{k}\right)=-\WH{P}^L_{\mu}\cdot
M^{\mu}\left(i+1,...,\WH{k}\right)=0 \eea

Because we have summed over all four polarization vectors, the
result (\ref{2.16}) does not depend on the gauge choice of $P$ and
we can choose gauge freely. The building block of (\ref{2.16}) is
three-point on-shell amplitude and $J^\mu(1,2)$. Without a loss of
generality, two-point off-shell currents are given as
%
\bea
J^{\mu}\left(1^{-},2^{+}\right) & = & \frac{1}{\sqrt{2}s_{12}}
\left(\frac{[r_{1}2]\langle1r_{2}\rangle}{[r_{1}1]\langle r_{2}2\rangle}
\left(1-2\right)^{\mu}+\frac{[2r_{1}]\langle21\rangle}{[r_{1}1]\langle r_{2}2
\rangle}\langle r_{2}|\gamma^{\mu}|2]+\frac{[12]\langle r_{2}1\rangle}{[r_{1}1]
\langle r_{2}2\rangle}[r_{1}|\gamma^{\mu}|1\rangle\right)\Label{3.17}\\
J^{\mu}\left(1^{+},2^{+}\right) & = &
\frac{1}{\sqrt{2}s_{12}}\left(\frac{[12]\langle r_{2}r_{1}
\rangle}{\langle r_{1}1\rangle\langle
r_{2}2\rangle}\left(1-2\right)^{\mu}+\frac{[21] \langle
r_{1}2\rangle}{\langle r_{1}1\rangle\langle r_{2}2\rangle}\langle
r_{2}|\gamma^{\mu}|2] +\frac{[21]\langle r_{2}1\rangle}{\langle
r_{1}1\rangle\langle r_{2}2\rangle}\langle r_{1}|
\gamma^{\mu}|1]\right)~~\Label{2.18}\\
J^{\mu}\left(1^{-},2^{T}\right) & = &\frac{1}{\sqrt2s_{12}}\left(-\frac{\Spaa{12}\Spbb{2r_1
}}{\Spbb{r_11}}\left(1+2\right)^{\mu}+\frac{\Spaa{12}\Spbb{21}}{\Spbb{r_11}}\Spab{1|\gamma^{\mu}|r_1}\right)
~~\Label{2.18-1}\\
J^{\mu}\left(1^{+},2^{T}\right) & =
&\frac{1}{\sqrt2s_{12}}\left(\frac{\Spaa
{r_12}\Spbb{21}}{\Spaa{r_11}}\cdot\left(1+2\right)^{\mu}
-\frac{\Spaa{12}\Spbb{21}}{\Spaa{r_11}}\Spab{r_1|\gamma^{\mu}|1}\right)~~\Label{2.18-2}
\eea
where the gauge of each on-shell gluon has kept. Having established
the general idea for recursion relation, we present two examples.

   \begin{figure}[hbt]
  \centering
  \scalebox{0.75}[0.75]{\includegraphics{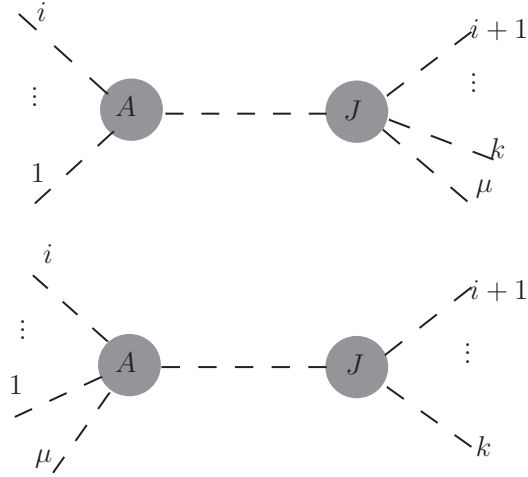}}\quad
  \caption{Two parts in the  recursion relation of off-shell gluon current \Label{Fig:division-22}}
\end{figure}

\subsection*{Example 1}

The first example is  three-point current
$J^{\mu}\left(1^{+},2^{+},3^{+}\right)$. With the  deformation on
$p_{1}$ and $p_{2}$,
\bea
|1]\rightarrow|1]-z|2],\,|2\rangle\rightarrow|2\rangle+z|1\rangle\eea
we can write down the recursion relation as
\bea
J^{\mu}\left(1^{+},2^{+},3^{+}\right)=J^{\mu}\left(\WH{1}^{+},\WH{P}^{+}\right)\cdot\frac{1}{s_{23}}\cdot
A\left(-\WH{P}^{-},\WH{2}^{+},3^{+}\right)\nonumber\\
+J^{\mu}\left(\WH{1}^{+},\WH{P}^{-}\right)\cdot\frac{1}{s_{23}}\cdot
A\left(-\WH{P}^{+},\WH{2}^{+},3^{+}\right)\nonumber\\
+J^{\mu}\left(\WH{1}^+,\WH{P}^L\right)\cdot\frac{1}{s_{23}}\cdot
A\left(-\WH{P}^T,\WH{2}^+,3^+\right)\nonumber\\
+J^{\mu}\left(\WH{1}^+,\WH{P}^T\right)\cdot\frac{1}{s_{23}}\cdot
A\left(-\WH{P}^L,\WH{2}^+,3^+\right)~~~\Label{J-Exa1-1}\eea
here $\WH{P}^L$ and $\WH{P}^T$ are longitude and timelike vectors of
the new gluon $\WH{P}$. We note here that as  the external gluons
are color-ordered, the order in the current is constrained which
leads to only one pole appear in the recursion relation above, i.e.,
there is no term $A(3^+, \WH 1^+, \WH P^{h}){1\over s_{13}} J^\mu
(-\WH P^{\W h}, 2)$.

For the four terms in (\ref{J-Exa1-1}),   the second term is zero
with all positive helicities and the fourth term is zero by Ward
Identity. Then the recursion relation is given only by
\bea
J^{\mu}\left(1^{+},2^{+},3^{+}\right)=J^{\mu}\left(\WH{1}^{+},\WH{P}^{+}\right)\cdot\frac{1}{s_{23}}\cdot
A\left(-\WH{P}^{-},\WH{2}^{+},3^{+}\right)+J^{\mu}\left(\WH{1}^+,\WH{P}^L\right)\cdot\frac{1}{s_{23}}\cdot
A\left(-\WH{P}^T,\WH{2}^+,3^+\right)\label{reducedrecursion}\eea
To check the $\WH P$-gauge independent of result we set the
reference momentum of the new gluon $\WH{P}$ to be an arbitrary null
vector $q$ and reference momenta of external particles to be
$r_{1}=r_{2}=r_{3}=r$, thus two terms are respectively given by
\bea
&&J^{\mu}\left(\WH{1}^{+},\WH{P}^{+}\right)\cdot\frac{1}{s_{23}}\cdot
A\left(-\WH{P}^{-},\WH{2}^{+},3^{+}\right)\nonumber\\
&&=\frac{1}{\sqrt{2}s_{\WH{1}\WH{P}}}\left(\frac{\Spbb{\WH{P}\WH{1}}\Spaa{rq}}{\Spaa{r\WH{1}}\Spaa{q\WH{P}}}\left(\WH{1}-\WH{P}\right)^{\mu}+
\frac{\Spbb{\WH{1}\WH{P}} \Spaa{\WH{P}r}}{\Spaa{r\WH{1}}\Spaa{q\WH{P}}}\Spab{
q|\gamma^{\mu}|\WH{P}} +\frac{\Spbb{\WH{1}\WH{P}}\Spaa{\WH{1}q}}{\Spaa{r\WH{1}}\Spaa{q\WH{P}}}\Spab{ r|
\gamma^{\mu}|\WH{1}}\right)\cdot\frac{1}{s_{23}}\cdot\frac{\Spbb{\WH{2}3}^3}{\Spbb{\WH{2}\WH{P}}\Spbb{\WH{P}3}}\;,\;\\
&&J^{\mu}\left(\WH{1}^+,\WH{P}^L\right)\cdot\frac{1}{s_{23}}\cdot
A\left(-\WH{P}^T,\WH{2}^+,3^+\right)\nonumber\\
&&=-\frac{1}{\sqrt2s_{\WH{1}\WH{P}}}\left(\frac{\Spaa
{r\WH{P}}[\WH{P}\WH{1}]}{\Spaa{r\WH{1}}}\cdot\left(\WH{1}+\WH{P}^L\right)^{\mu}
-\frac{\Spaa{\WH{1}\WH{P}}[\WH{P}\WH{1}]}{\Spaa{r\WH{1}}}\Spab{r|\gamma^{\mu}|\WH{1}}\right)
\cdot\frac{1}{s_{23}}\cdot\frac{\Spaa{qr}\Spbb{\WH{2}3}\Spaa{r\WH{P}}}{\Spaa{q\WH{P}}\Spaa{r3}\Spaa{r\WH{2}}}
~~~\Label{gauge-extra-1}\eea
There are several things we want to discuss regarding this result.
First the $q$-gauge independent can be numerically checked using the
package S@M \cite{Maitre:2007jq}
 and indeed it is given by
 \bea
J^{\mu}\left(1^{+},2^{+},3^{+}\right)&=&J^{\mu}\left(\WH{1}^{+},\WH{P}^{+}\right)\cdot\frac{1}{s_{23}}\cdot
A\left(-\WH{P}^{-},\WH{2}^{+},3^{+}\right)+J^{\mu}\left(\WH{1}^+,\WH{P}^L\right)\cdot\frac{1}{s_{23}}\cdot
A\left(-\WH{P}^T,\WH{2}^+,3^+\right)\nonumber\\
&=&\frac{\langle r|\gamma^{\mu}\slashed{k}_{123}|r\rangle}{\sqrt{2}\langle r1
\rangle\langle12\rangle\langle23\rangle\langle3r\rangle}\eea
as we expected. We have seen that to achieve the $q$-gauge
independent, the second term is very crucial with the unfamiliar
$A\left(-\WH{P}^T,\WH{2}^+,3^+\right)$. In particular, the gauge
choice of gluon $2,3$ will effect the whole result through
$A\left(-\WH{P}^T,\WH{2}^+,3^+\right)$, which is not gauge
invariant.


Secondly,  it's very obvious that with $q=r$, the second term
(\ref{gauge-extra-1}) vanishes, thus the result is given just by
familiar on-shell BCFW recursion relation for amplitude
 \bea
J^{\mu}\left(1^{+},2^{+},3^{+}\right) & = & J^{\mu}\left(\WH{1}^{+},\WH{P}^{+}\right)\cdot\frac{1}{s_{23}}\cdot A\left(-\WH{P}^{-},\WH{2}^{+},3^{+}\right)\nonumber\\
 & = & \frac{1}{\sqrt{2}\langle1\WH{P}\rangle}\left(\frac{\langle r|\gamma^{\mu}|\WH{P}]}{\langle r1\rangle}+\frac{\langle r|\gamma^{\mu}|\WH{1}]}{\langle r\WH{P}\rangle}\right)\cdot\frac{1}{s_{23}}\cdot-\frac{[\WH{2}3]}{[3\WH{P}][\WH{P}\WH{2}]}\nonumber\\
 & = & \frac{\langle r|\gamma^{\mu}\slashed{k}_{123}|r\rangle}
 {\sqrt{2}\langle r1\rangle\langle12\rangle\langle23\rangle\langle3r\rangle}\eea
We must emphasize this is true when and only when we choose the
special gauge.

\subsection*{Example 2}
Using the same method to the current
$J^{\mu}\left(1^{-},2^{+},3^{+}\right)$, with the same deformation\[
|1]\rightarrow|1]-z|2],\,|2\rangle\rightarrow|2\rangle+z|1\rangle\]
the recursion relation is given as 
\bea
J^{\mu}\left(1^{-},2^{+},3^{+}\right)  =  J^{\mu}\left(\WH{1}^{-},\WH{P}^{+}\right)\cdot\frac{1}{s_{23}}\cdot A\left(-\WH{P}^{-},\WH{2}^{+},3^{+}\right)\nonumber\\
 +J^{\mu}\left(\WH{1}^{-},\WH{P}^{-}\right)\cdot\frac{1}{s_{23}}\cdot A\left(-\WH{P}^{+},\WH{2}^{+},3^{+}\right)\nonumber\\
 +J^{\mu}\left(\WH{1}^{-},\WH{P}^{L}\right)\cdot\frac{1}{s_{23}}\cdot A\left(-\WH{P}^{T},\WH{2}^{+},3^{+}\right)\nonumber\\
 +J^{\mu}\left(\WH{1}^{-},\WH{P}^{T}\right)\cdot\frac{1}{s_{23}}\cdot A\left(-\WH{P}^{L},\WH{2}^{+},3^{+}\right)
\Label{3.23}\eea
where again color-ordering leads to only one cut $s_{23}$ as above.
In (\ref{3.23})  the second vanishes with all positive helicity
while the fourth term vanishes by Ward Identity. With general
reference null momentum of the new gluon $\WH{P}$, the first and
third terms are given as
\bea
 &&J^{\mu}\left(\WH{1}^{-},\WH{P}^{+}\right)\cdot\frac{1}{s_{23}}\cdot
 A\left(-\WH{P}^{-},\WH{2}^{+},3^{+}\right)\nonumber\\
&&=\frac{1}{\sqrt2s_{\WH{1}\WH{P}}}\left[-\frac{\Spaa{\WH{1}q}\Spbb{\WH{P}\WH{2}}}
{\Spbb{\WH{2}\WH{1}}\Spaa{q\WH{P}}}\left(\WH{1}-\WH{P}\right)^{\mu}-
\frac{\Spaa{\WH{P}\WH{1}}\Spbb{\WH{2}\WH{P}}}
{\Spaa{q\WH{P}}\Spbb{\WH{2}\WH{1}}}\Spab{q|\gamma^{\mu}
|\WH{P}}+\frac{\Spaa{\WH{1}q}\Spbb{\WH{P}\WH{1}}}
{\Spbb{\WH{2}\WH{1}}\Spaa{q\WH{P}}}\Spab{\WH{1}|\gamma^{\mu}|
\WH{2}}\right]\cdot\frac{1}{s_{23}}\cdot\frac{\Spbb{\WH{2}3}^3}
{\Spbb{\WH{2}\WH{P}}\Spbb{\WH{P}3}}\nonumber\\
&&J^{\mu}\left(\WH{1}^{-},\WH{P}^{L}\right)\cdot\frac{1}{s_{23}}\cdot A\left(-\WH{P}^{T},\WH{2}^{+},3^{+}\right)\nonumber\\
&&=\frac{1}{\sqrt2s_{\WH{1}\WH{P}}}\left[-\frac{\Spaa{\WH{1}\WH{P}}\Spbb{\WH{P}\WH{2}
}}{\Spbb{\WH{2}\WH{1}}}\left(\WH{1}+\WH{P}\right)^{\mu}+\frac{\Spaa{\WH{1}
\WH{P}}\Spbb{\WH{P}\WH{1}}}{\Spbb{\WH{2}\WH{1}}}\Spab{\WH{1}|\gamma^{\mu}|\WH{2}}\right]\cdot\frac{1}{s_{23}}\cdot\frac{\Spaa{\WH{P}\WH{1}}\Spaa{q\WH{1}}\Spbb{\WH{2}3}}{\Spaa{\WH{P}q}\Spaa{\WH{1}\WH{2}}\Spaa{\WH{1}3}}
~~~\Label{1.33-1.33}\eea
and it is  numerically checked that the result is $q$-gauge
invariant. Recall that by (\ref{3.17}), a good gauge choosing of
current $J^{\mu}\left(1^{-},2^{+}, 3^{+}\right)$ is $r_1=p_2$,
$r_2=r_3=p_1$. Also by checking (\ref{1.33-1.33}) it is easy to see
that when we choose $q=p_1$, many terms will be zero. Putting this
choice back we get immediately
\bea
J^{\mu}\left(1^{-},2^{+},3^{+}\right)=\frac{\Spbb{32}\Spaa{1|\gamma^{\mu}\slashed{k}_{123}|1}}
{\sqrt{2}s_{12}s_{123}\Spaa{23}} \eea
Again, we find the results
from on-shell recursion relation and off-shell recursion relation
match with each other.

From above two examples, it is easy to see that although with
off-shell current, which is not gauge invariant, we need to sum over
four helicity configurations in recursion relation, there is gauge
freedom of $\WH P$ we can choose to eliminate many middle
contributions. Properly using of this observation will simplify
calculations.

\subsection{On shell recursion relation involving the off shell leg}
Through derivation above, we show that BCFW recursion relation is
valid for the gluon current with deformation of two on-shell
particles.
However as a gluon current contains an off shell leg, there seems to
be another deformation we can make  such that $z$-dependent momentum
flux goes through the current from an on-shell particle to the off
shell leg. In this part we will exhibit how this could be realized
and what's the recursion relation it will imply.

To find the recursion relation involving the off shell leg in a
current, we consider an one-particle shifting deformation. Without
loss of generality, we assume the first gluon in the current has $+$
helicity. For such a current $J^{\mu}(1^+,2,...,k)$, we do the
deformation as
\bea
|1\rangle\rightarrow|1\rangle+z|q\rangle\label{oneshift+}\eea
where $|q\rangle$ is the left-handed spinor of a arbitrary lightlike
momentum $q$. At this momentum, it seems that $q$ can be chosen
arbitrarily, but from explicit results, for example (\ref{2.6}), we
can see that there is unphysical pole, for example $\Spaa{r|1}$,
shown up. To get rid of this phenomenon  and keep only physical
pole, we should choose $q$ to be the same gauge choice for the
definition of positive helicity of particle $1$.

This deformation has kept the on-shell condition for particle $1$,
and there is no requirement of momentum conservation because the
off-shell momentum is allowed to change. With this deformation, the
polarization vector will behave as
\bea \epsilon_1^{+\mu}=\frac{\langle
r_1|\gamma^{\mu}|p_1]}{\sqrt2\langle
r_1p_1\rangle}\sim\frac{1}{z}\eea
for large $z$. To consider the large $z$-behavior, we consider the
path from $1$ to off-shell leg with only most dangerous cubic
vertexes, since  each propagator contributes $\frac{1}{z}$ and each
cubic vertex contributes $z$, the overall  $z$ behavior will be
$\frac{1}{z}\cdot
z^i\cdot\frac{1}{z^i}=\frac{1}{z}$.~~~\footnote{For the case that
the first gluon has $-$ helicity, the deformation should be $
|1]\rightarrow|1]-z|q]$ so the polarization will behave as $
\epsilon_1^{-\mu}=-\frac{[r_1|\gamma^{\mu}|p_1\rangle}{\sqrt2[r_1p_1]}\sim\frac{1}{z}$.}

With the good behavior, the recursion relation is given as
\bea
J^{\mu}(1,2,...,k)=\sum_i\sum_{h,\tilde{h}}A(\WH{1},2,...,i-1,\WH{P}^h)\cdot\frac{1}{P_{1,i-1}^2}\cdot
J^{\mu}(-\WH{P}^{\tilde{h}},i,...,k),~~ ~~\Label{3.28} \eea
where the sum is over  $ (h,\W h)=(+,-),(-,+),(L,T),(T,L)$ for exact
same reason as in previous subsection. Different from the recursion
relation in (\ref{2.16}), there is only one term and only three
shifted momenta $\WH{1},\WH{P}^h,-\WH{P}^{-\tilde{h}}$ instead of
four. 
Also the off-shell momentum will be $z$-dependent, thus we will have
following $z$-dependent propagator
\bea D^{\mu\nu}(p-zl)=\frac{-ig^{\mu\nu}}{P^2-2zl\cdot
p},~~~l=\ket{q}\bket{1}\eea
in Feynman gauge, which will contribute to the residue. Finally
because the color ordering, the deformation with $1$ will give
minimum number of terms, but we could choose arbitrary on-shell
particle, which will be discussed in  an example.


Having established (\ref{3.28}) we give some examples.

   \begin{figure}[hbt]
  \centering
  \scalebox{0.75}[0.75]{\includegraphics{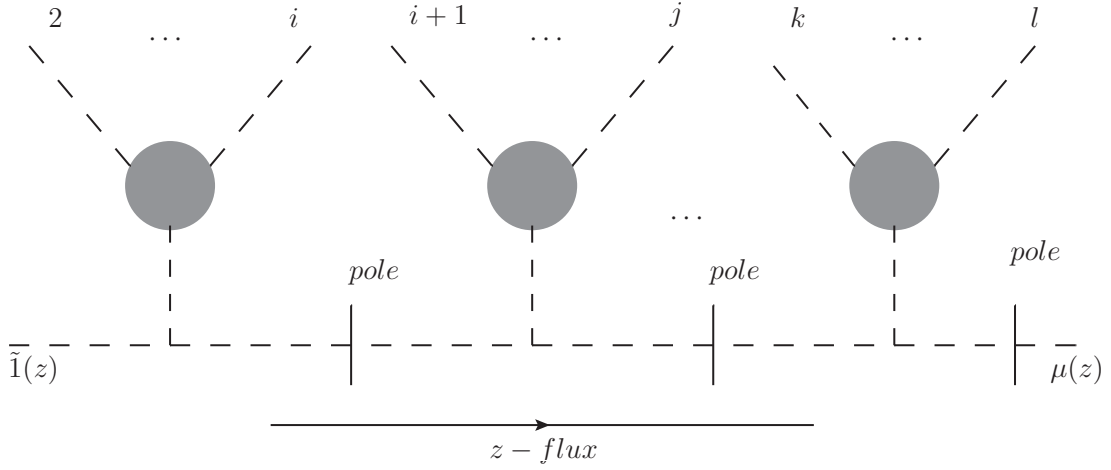}}\quad
  \caption{The $z$-flux in the gluon current encountering all 3-vertices \label{Fig:division-1}}
\end{figure}

\subsubsection*{Example 1 $J^{\mu}(1,2)\rightarrow J^{\mu}(1)$}

Let us start with the two-point current $J^{\mu}(1,2)$ which is the
simplest example. As for $J^{\mu}(1^+,2^+)$, a usual gauge choosing
is $r_1=r_2=r$, so we take the deformation as
\bea
|1\rangle\rightarrow|1\rangle+z|r\rangle\label{3.33}
\eea
to avoid unwanted unphysical pole. The recursion relation is given
by
\bea
J^{\mu}(1^+,2^+)=A(\WH{1}^+,2^+,\WH{P}^-)\cdot\frac{1}{s_{12}}\cdot J^{\mu}(-\WH{P}^+)+A(\WH{1}^+,2^+,\WH{P}^T)\cdot\frac{1}{s_{12}}\cdot J^{\mu}(-\WH{P}^L)
\eea
where another two helicity configurations are zero.  Without gauge
choosing of $P$, the result is
 \bea
J^{\mu}\left(1^+,2^+\right)=\frac{\Spbb{\WH{1}2}^3}
{\Spbb{\WH{1}\WH{P}}\Spbb{\WH{P}2}}\cdot\frac{1}
{s_{12}}\cdot\frac{\Spab{q|\gamma^{\mu}|\WH{P}}}{\sqrt2\Spaa{q\WH{P}}}
+\frac{\sqrt2\Spaa{qr}\Spbb{\WH{1}2}\Spaa{r\WH{P}}}
{\Spaa{q\WH{P}}\Spaa{r\WH{1}}\Spaa{r2}}\cdot\frac{1}{s_{12}}
\cdot\frac{\Spab{\WH{P}|\gamma^{\mu}|\WH{P}}}{2}\eea
which can be checked to be $q$-gauge independent by Mathematica. To
simplify analytically,  we can choose the convenient gauge
$q_{\WH{P}}=r$,  so the second term vanishes and the result is
\bea
J^{\mu}(1^+,2^+)&=&-\frac{\Spbb{12}^3}{\Spbb{2\WH{P}}\Spbb{\WH{P}1}}
\cdot\frac{1}{s_{12}}\cdot\frac{\Spab{r|\gamma^{\mu}|\WH{P}}}{\sqrt2\Spaa{r\WH{P}}}
=\frac{\Spaa{r|\gamma^{\mu}\slashed{P}_{12}|r}}{\sqrt2\Spaa{r1}\Spaa{12}\Spaa{2r}}
\eea

There is one technical issue with the choice $q=r$. Naively  the
deformed polarization vector behaves as $\epsilon_1^{\mu}\sim z^0$
which seems to destroy the good large $z$ behavior. However the
first cubic vertex connecting $1$ and $2$ now is also behaves as
$z^0$ instead of $z^1$
\bea
V_3(z)&=&\frac{i}{\sqrt2}\left[\eta^{\nu\rho}\cdot(p_1-
p_2)^{\mu}+2\eta^{\rho\mu}\cdot p_2^{\nu}-2\eta^{\mu\nu}\cdot p_1^
{\rho}\right]\cdot\epsilon_{1\nu}\cdot\epsilon_{2\rho}\nonumber\\
&=&\sqrt2i\epsilon_2^{\mu}\cdot (p_2\cdot\epsilon_1)\sim\ z^0~, \eea
thus the whole $z$ behavior is still $\frac{1}{z}$ and the recursion relation is still valid.

\subsubsection*{Example 2 $J^{\mu}(1,2,3)\rightarrow J^{\mu}(1,2)$}

For this example, we will consider different choices of
deformations.

\subsubsection*{a. $J^{\mu}(1^+,2^+,3^+)$ with with deformation on $1^+$}

With deformation (\ref{3.33}) there are  two poles in the current so
the recursion relation is given as
\bea
J^{\mu}(1^+,2^+,3^+)=A(\WH{1}^+,2^+,\WH{P}^-)\cdot\frac{1}{s_{12}}\cdot J^{\mu}(-\WH{P}^+,3^+)\nonumber\\
+A(\WH{1}^+,2^+,\WH{P}^T)\cdot\frac{1}{s_{12}}\cdot J^{\mu}(-\WH{P}^L,3^+)
\nonumber\\
+A(\WH{1}^+,2^+,3^+,\WH{P}^T)\cdot\frac{1}{s_{123}}\cdot J^{\mu}(-\WH{P}^L)
\eea
where among eight possible contributions we have kept only  three
with nonzero contributions. With general reference momentum $q$ of
the new gluon $\WH{P}$, these three terms are given by
\bea
&&A(\WH{1}^+,2^+,\WH{P}^-)\cdot\frac{1}{s_{12}}\cdot J^{\mu}(-\WH{P}^+,3^+)\nonumber\\&&=-\frac{\Spbb{\WH{1}2}^3}{\Spbb{2\WH{P}}\Spbb{\WH{P}\WH{1}}}\cdot\frac{1}{s_{12}}\cdot\frac{1}{\sqrt2s_{3\WH{P}}}\left(\frac{\Spbb{\WH{P}3}\Spaa{rq}}{\Spaa{q\WH{P}}\Spaa{r3}}\left(\WH{P}-3\right)^{\mu}+\frac{\Spbb{3
\WH{P}}\Spaa{q3}}{\Spaa{q\WH{P}}\Spaa{r3}}\Spab{r|\gamma^{\mu}|3}+\frac{\Spbb{3\WH{P}}\Spaa{r\WH{P}}}{\Spaa{q\WH{P}}\Spaa{r3}}\Spaa{q|\gamma^{\mu}|\WH{P}}\right)\\
&&A(\WH{1}^+,2^+,\WH{P}^T)\cdot\frac{1}{s_{12}}\cdot J^{\mu}(-\WH{P}^L,3^+)\nonumber\\
&&=\frac{\Spbb{\WH{1}2}\Spaa{qr}\Spaa{\WH{P}r}}{\Spaa{r\WH{1}}\Spaa{2r}\Spaa{\WH{P}q}}\cdot\frac{1}{s_{12}}\cdot\frac{1}{\sqrt2s_{3\WH{P}}}\left(\frac{\Spaa{r\WH{P}}\Spbb{\WH{P}3}}{\Spaa{r3}}\left(3+\WH{P}\right)^{\mu}-\frac{\Spaa{3\WH{P}}\Spbb{\WH{P}3}}
{r3}\Spab{r|\gamma^{\mu}|3}\right)\\
&&A(\WH{1}^+,2^+,3^+,\WH{P}^T)\cdot\frac{1}{s_{123}}\cdot J^{\mu}(-\WH{P}^L)\nonumber\\
&&=\frac{1}{\sqrt2}\cdot\frac{\Spaa{rq}\Spaa{\WH{P}r}}{\Spaa{r\WH{1}}\Spaa{\WH{1}2}\Spaa{23}\Spaa{3r}}\cdot\Spab{\WH{P}|\gamma^{\mu}|\WH{P}}
\eea
and  we have checked that the sum is same for any choice of $q$. We
can simplify  result by  choosing the gauge of $P$ to be
$r_{\WH{P}}=r_1=r_2=r_3=r$ and again, the second and third terms
vanish  with factor  $\Spaa{qr}$. Finally  we have
\bea
J^{\mu}(1^+,2^+,3^+)&=&-\frac{\Spbb{12}^3}{\Spbb{2\WH{P}}\Spbb{\WH{P}1}}\cdot\frac{1}{s_{12}}\cdot\frac{1}{\sqrt2s_{3\WH{P}}}\left(\frac{\Spbb{3\WH{P}}}{\Spaa{r\WH{P}}}\Spab{r|\gamma^{\mu}|3}+\frac{[3\WH{P}]}{\Spaa{r3}}\Spab{r|\gamma^{\mu}|\WH{P}}\right)\nonumber\\
&=&\frac{\Spaa{r|\gamma^{\mu}\slashed{P}_{123}|r}}{\sqrt2\Spaa{r1}\Spaa{12} \Spaa{23} \Spaa{3r}}
\eea
which gives the right result.

\subsubsection*{b. $J^{\mu}(1^+,2^+,3^+)$ with deformation on $2^+$}
If we consider the same current $J^{\mu}(1^+,2^+,3^+)$ with
deformation \bea |2\rangle\rightarrow|2\rangle+z|r\rangle\eea the
recursion relation then will contain different poles. Among many
terms, there with nonzero contributions are
\bea
J^{\mu}(1^+,2^+,3^+)=&&A(1^+,\WH{2}^+,\WH{P}^-)\cdot\frac{1}{s_{12}}\cdot
J^{\mu}(-\WH{P}^+,3^+)
+A(1^+,\WH{2}^+,\WH{P}^T)\cdot\frac{1}{s_{12}}\cdot J^{\mu}(-\WH{P}^L,3^+)\nonumber\\
&&+J^{\mu}(1^+,-\WH{P}^+)\cdot\frac{1}{s_{23}}\cdot A(\WH{P}^-,\WH{2}^+,3^+)
+J^{\mu}(1^+,-\WH{P}^L)\cdot\frac{1}{s_{23}}\cdot A(\WH{P}^T,\WH{2}^+,3^+)\nonumber \\
&&+A(1^+,\WH{2}^+,3^+,\WH{P}^T)\cdot\frac{1}{s_{123}}\cdot
J^{\mu}(-\WH{P}^L) \eea
with following explicit expressions\footnote{In principle, the
reference momenta for $\WH P_{12}, \WH P_{23}, \WH P_{123}$ can be
different. Here for simplicity we have chosen them to be same.}
\bea
&&A(1^+,\WH{2}^+,\WH{P}^-)\cdot\frac{1}{s_{12}}\cdot J^{\mu}(-\WH{P}^+,3^+)\nonumber\\
&&=-\frac{\Spbb{1\WH{2}}^3}{\Spbb{\WH{2}\WH{P}}\Spbb{\WH{P}1}}\cdot\frac{1}{s_{12}}\cdot\frac{1}{\sqrt2s_{3\WH{P}}}\cdot\left(\frac{\Spbb{\WH{P}3}\Spaa{rq}}{\Spaa{q\WH{P}}\Spaa{r3}}\left(\WH{P}-3\right)^{\mu}+\frac{\Spbb{3
\WH{P}}\Spaa{q3}}{\Spaa{q\WH{P}}\Spaa{r3}}\Spab{r|\gamma^{\mu}|3}+\frac{\Spbb{3\WH{P}}\Spaa{r\WH{P}}}{\Spaa{q\WH{P}}\Spaa{r3}}\Spab{q|\gamma^{\mu}|\WH{P}}\right)\\
&&A(1^+,\WH{2}^+,\WH{P}^T)\cdot\frac{1}{s_{12}}\cdot J^{\mu}(-\WH{P}^L,3^+)\nonumber\\
&&=\frac{\Spbb{1\WH{2}}\Spaa{qr}\Spaa{\WH{P}r}}{\Spaa{r1}\Spaa{\WH{2}r}\Spaa{\WH{P}q}}\cdot\frac{1}{s_{12}}\cdot\frac{1}{\sqrt2s_{3\WH{P}}}\left(\frac{\Spaa{r\WH{P}}\Spbb{\WH{P}3}}{\Spaa{r3}}\left(3+\WH{P}\right)^{\mu}-\frac{\Spaa{3\WH{P}}\Spbb{
\WH{P}3}}{\Spaa{r3}}\Spab{r|\gamma^{\mu}|3}\right)\\
&&J^{\mu}(1^+,-\WH{P}^+)\cdot\frac{1}{s_{23}}\cdot A(\WH{P}^-,\WH{2}^+,3^+)\nonumber\\
&&=\frac{1}{\sqrt2s_{1\WH{P}}}\left(\frac{\Spbb{1\WH{P}}\Spaa{rq}}{\Spaa{r1}\Spaa{q\WH{P}}}\left(1-\WH{P}\right)^{\mu}+\frac{\Spbb{
\WH{P}1}\Spaa{rq}}{\Spaa{r1}\Spaa{q\WH{P}}}\Spab{q|\gamma^{\mu}|\WH{P}}+\frac{\Spbb{\WH{P}1}\Spaa{q1}}{\Spaa{r1}\Spaa{q\WH{P}}}\Spab{r|\gamma^{\mu}|1}\right)\cdot\frac{1}{s_{23}}\cdot\frac{-\Spbb{\WH{2}3}^3}{\Spbb{3\WH{P}}\Spbb{\WH{P}2}}\\
&&J^{\mu}(1^+,-\WH{P}^L)\cdot\frac{1}{s_{23}}\cdot A(\WH{P}^T,\WH{2}^+,3^+)\nonumber\\
&&=\frac{\Spbb{\WH{2}3}\Spaa{qr}\Spaa{\WH{P}r}}{\Spaa{r\WH{2}}\Spaa{3r}\Spaa{\WH{P}q}}\cdot\frac{1}{s_{23}}\cdot\frac{1}{\sqrt2s_{1\WH{P}}}\left(\frac{\Spaa{r\WH{P}}\Spbb{\WH{P}1}}{\Spaa{r1}}\left(1+\WH{P}\right)^{\mu}-\frac{\Spaa{1\WH{P}}\Spbb{
\WH{P}1}}{\Spaa{r1}}\Spab{r|\gamma^{\mu}|1}\right)\\
&&A(1^+,\WH{2}^+,3^+,\WH{P}^T)\cdot\frac{1}{s_{123}}\cdot J^{\mu}(-\WH{P}^L)\nonumber\\
&&=\frac{1}{\sqrt2}\cdot\frac{\Spaa{rq}\Spaa{\WH{P}r}}{\Spaa{r1}\Spaa{1\WH{2}}\Spaa{\WH{2}3}\Spaa{3r}}\cdot\Spab{\WH{P}|\gamma^{\mu}|\WH{P}}
\eea
Now we choose the good gauge $q_{\WH P_{23}}=q_{\WH P_{12}}=q_{\WH
P_{123}}=r$, so the second and the forth terms vanish and the others
are given respectively as
\bea
A(1^+,\WH{2}^+,\WH{P}^-)\cdot\frac{1}{s_{12}}\cdot J^{\mu}(-\WH{P}^+,3^+)&=&-\frac{\langle r|\gamma^{\mu}\slashed{P}_{123}|r\rangle}{\langle r2\rangle\langle 12\rangle\langle 31\rangle\langle 3r\rangle}\nonumber\\
J^{\mu}(1^+,-\WH{P}^+)\cdot\frac{1}{s_{23}}\cdot A(\WH{P}^-,\WH{2}^+,3^+)&=&-\frac{\langle r|
\gamma^{\mu}\slashed{P}_{123}|r\rangle}{\langle r1\rangle\langle 23\rangle\langle 13\rangle\langle r2\rangle}\eea
Adding them together  we get the wanted result
\bea
J^{\mu}(1^+,2^+,3^+)=\frac{\langle
r|\gamma^{\mu}\slashed{P}_{123}|r\rangle}{\sqrt2\langle
r1\rangle\langle 12\rangle \langle 23\rangle \langle 3r\rangle}\eea

\subsubsection*{c. $J^{\mu}(1^-,2^+,3^+)$}

For a current with all $+$ helicity, we usually choose  the
reference momenta to be $r_1=r_2=\cdots=r_k=r$. And through analysis
above we find this gauge choosing lead us to take $q_{\WH P}=r$ in
(\ref{oneshift+}) naturally. However, with current
$J^{\mu}(1^-,2^+,3^+)$ the gauge choosing is no longer the same.
Usually we choose $r_1=p_2$ and $r_2=r_3=p_1$, so good choice of $q$
for $\WH P$ will be different.

As for the first gluon has minus helicity now, we take the
deformation on right-handed spinor this time \bea
|1]\rightarrow|1]-z|\omega]\eea where as we have remarked before, to
avoid the spurious pole, we should set $\omega=r_1$. However, at
this moment, we will leave $\omega$ undetermined. The recursion relation is given by
%
\bea
J^{\mu}(1^-,2^+,3^+)=&&A(\WH{1}^-,2^+,\WH{P}^-)\cdot\frac{1}{s_{12}}\cdot J^{\mu}(-\WH{P}^+,3^+)+A(\WH{1}^-,2^+,\WH{P}^+)\cdot\frac{1}{s_{12}}\cdot J^{\mu}(-\WH{P}^-,3^+)\nonumber\\&&+A(\WH{1}^-,2^+,\WH{P}^T)\cdot\frac{1}{s_{12}}\cdot J^{\mu}(-\WH{P}^L,3^+)
+A(\WH{1}^-,2^+,3^+,\WH{P}^-)\cdot\frac{1}{s_{123}}\cdot J^{\mu}(-\WH{P}^+)\nonumber\\&&+A(\WH{1}^-,2^+,3^+,\WH{P}^T)\cdot\frac{1}{s_{123}}\cdot J^{\mu}(-\WH{P}^L)
\eea
where we have kept only nonzero terms. Expressions for these
terms\footnote{In principle, the reference momenta for $\WH P_{12},
 \WH P_{123}$ can be different. Here for simplicity we
have chosen them to be same.} are given as
\bea
&&A(\WH{1}^-,2^+,\WH{P}^-)\cdot\frac{1}{s_{12}}\cdot J^{\mu}(-\WH{P}^+,3^+)\nonumber\\
&&=\frac{\Spaa{\WH{P}\WH{1}}^3}{\Spaa{\WH{1}2}\Spaa{2\WH{P}}}\cdot\frac{1}{s_{12}}\cdot\frac{1}{\sqrt2s_{3\WH{P}}}\left(\frac{\Spbb{\WH{P}3}\Spaa{rq}}{\Spaa{q\WH{P}}\Spaa{r3}}\left(\WH{P}-3\right)^{\mu}+\frac{\Spbb{3
\WH{P}}\Spaa{q3}}{\Spaa{q\WH{P}}\Spaa{r3}}\Spab{r|\gamma^{\mu}|3}+\frac{\Spbb{3\WH{P}}\Spaa{r\WH{P}}}{\Spaa{q\WH{P}}\Spaa{r3}}\Spaa{q|\gamma^{\mu}|\WH{P}}\right)\\
&&A(\WH{1}^-,2^+,\WH{P}^+)\cdot\frac{1}{s_{12}}\cdot J^{\mu}(-\WH{P}^-,3^+)\nonumber\\
&&=-\frac{\Spbb{2\WH{P}}^3}{\Spbb{\WH{P}\WH{1}}\Spbb{\WH{1}2}}\cdot\frac{1}{s_{12}}\cdot\frac{1}{\sqrt2s_{3\WH{P}}}\left(\frac{\Spbb{q3}\Spaa{\WH{P}\WH{1}}}{\Spbb{q\WH{P}}\Spaa{\WH{1}3}}\left(\WH{P}+3\right)^{\mu}-\frac{
\Spbb{3q}\Spaa{3\WH{P}}}{\Spbb{q\WH{P}}\Spaa{\WH{1}3}}\Spab{\WH{1}|\gamma^{\mu}|3}+\frac{\Spbb{\WH{P}3}\Spaa{\WH{1}}}{\Spbb{q\WH{P}}\Spaa{\WH{1}3}}\Spba{q|\gamma^{\mu}|\WH{P}}\right)\\
&&A(\WH{1}^-,2^+,\WH{P}^T)\cdot\frac{1}{s_{12}}\cdot J^{\mu}(-\WH{P}^L,3^+)=0\\
&&A(\WH{1}^-,2^+,3^+,\WH{P}^-)\cdot\frac{1}{s_{123}}\cdot J^{\mu}(-\WH{P}^+)\nonumber\\
&&=\frac{\Spaa{\WH{1}\WH{P}}^4}{\Spaa{\WH{1}2}\Spaa{23}\Spaa{3\WH{P}}\Spaa{\WH{P}\WH{1}}}\cdot
\frac{1}{s_{123}}\cdot\frac{\Spab{q|\gamma^{\mu}|\WH{P}}}{\sqrt2\Spaa{q\WH{P}}}\\
&&A(\WH{1}^-,2^+,3^+,\WH{P}^T)\cdot\frac{1}{s_{123}}\cdot J^{\mu}(-\WH{P}^L)\nonumber\\
&&=-\frac{1}{\sqrt2s_{12}s_{123}}\cdot\frac{\Spbb{32}\Spaa{\WH{1}q}\Spaa{\WH{P}
\WH{1}}}{\Spaa{23}\Spaa{q\WH{P}}}\Spab{\WH{P}|\gamma^{\mu}|\WH{P}}
\eea
and it can be checked that  the sum is equal to the off-shell
calculation with any choice of $q$. Now we put the $\omega=r_1=p_2$
back, then $\Spaa{\WH{P}1}=\Spbb{2\WH{P}}=0 $,  thus the first two
terms vanish and only the third term remains which gives
\bea
J^{\mu}\left(1^-,2^+,3^+\right)&=&A(\WH{1}^-,2^+,3^+,\WH{P}^-)\cdot\frac{1}{s_{123}}\cdot J^{\mu}\left(-\WH{P}^+\right)\nonumber\\
&=&\frac{\Spab{1|\slashed{P}_{23}|q}}{\Spaa{12}\Spaa{23}\Spab{3|\slashed{P}_{12}|q}}\cdot\frac{1}{s_{123}}\cdot\frac{\Spaa{1|\gamma^{\mu}\slashed{P}_{123}|1}}{\sqrt2}\nonumber\\
&=&\frac{\Spaa{1|\gamma^{\mu}\slashed{P}_{123}|1}}{\sqrt2\Spaa{12}\Spaa{23}\Spaa{31}}\frac{\Spaa{1|\slashed{3}\slashed{P}_{123}|1}}{s_{12}s_{123}}
\eea
and  is exactly the result from off shell calculation.

With these three examples, we show that the one particle  shifting
recursion relation is not only valid but also practical. One thing
we want to emphasize is that the shifted spinor should be same as
the one defined the corresponding helicity to cancel  the unphysical
poles shown up in the expression of current.


\section{The boundary contribution with fermion deformation in QCD }\label{boundaryQCD}

One motivation of our study is to understand  boundary contributions
in various situations. From previous studies, it has been found that
the difficulty of analysis increases with complexity of wave
functions of external particles. In this section we will consider
possible boundary contributions from deformation of two massless
fermions. To be more concretely, the example will be the process
$q\bar{q}\rightarrow ng$ in QCD, although it is well known
\cite{ArkaniHamed:2008yf, Cheung:2008dn} that there is a good
deformation of two gluons without boundary contributions.

Let us start with analyzing  the behavior of $A(z\to \infty)$.
Because fermions are massless,  there are only two possible helicity configurations
$A\left(q^{-},\bar{q}^{+},g_{1},g_{2},...,g_{n}\right)$ and
$A\left(q^{+},\bar{q}^{-},g_{1},g_{2},...,g_{n}\right)$. For
$A\left(q^{-},\bar{q}^{+},g_{1},g_{2},...,g_{n}\right)$, using the
Feynman rule we can see the general pattern of expressions is
$\Spab{q|...|\overline{ q}}$ while for
$A\left(q^{+},\bar{q}^{-},g_{1},g_{2},...,g_{n}\right)$, it is
$\Spba{q|...|\overline q}$. Thus if we take the deformation
\begin{equation}
|q]\rightarrow|q]-z|\bar{q}],\quad |\bar{q}\rangle\rightarrow
|\bar{q}\rangle+ z|{q}\rangle~~\Label{3.1}\end{equation}
there will be $z^0$ from wave function for
$A\left(q^{-},\bar{q}^{+},g_{1},g_{2},...,g_{n}\right)$ or $z^2$ for
$A\left(q^{+},\bar{q}^{-},g_{1},g_{2},...,g_{n}\right)$. Since the
$z$-dependence flows along the fermion line, we can see that the
vertex does not depend on $z$ and the fermion propagator
$\frac{i\slashed{P}}{P^2}$ gives overall ${z\over z}\sim z^0$. Thus
the large $z$-behavior will be $A(z)\to z^0$ or $A(z)\to z^2$. To
make the problem simpler, we will take the deformation such that the
large $z$-behavior is  $A(z)\to z^0$, i.e., for
$A\left(q^{+},\bar{q}^{-},g_{1},g_{2},...,g_{n}\right)$, we should
exchange the role of $q$ and $\bar{q}$ in (\ref{3.1}).

\begin{figure}[hbt]
  \centering
  \scalebox{1.13}[1.13]{\includegraphics{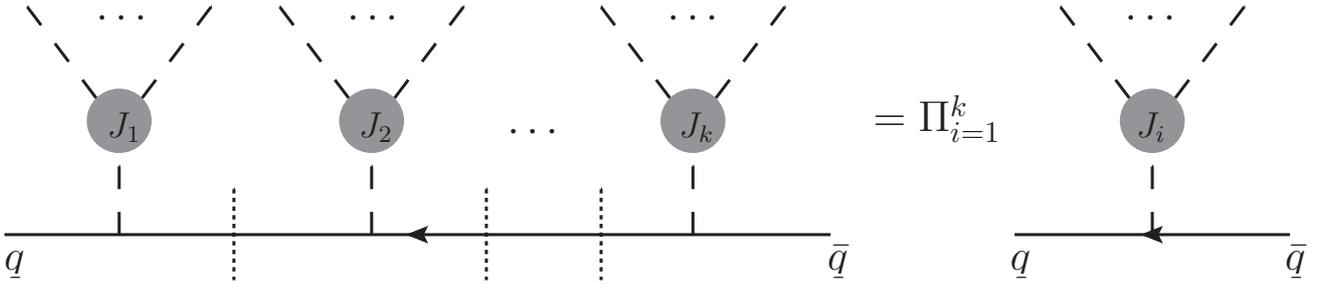}}\quad
  \caption{A graphic description for the boundary term.  \label{Fig:division}}
\end{figure}

Now we will work out the boundary contributions for  amplitude
$A\left(q^{-},\bar{q}^{+},g_{1},g_{2},...g_{n}\right)$. The general
expressions  of Feynman diagrams could be written as
\begin{equation} F=\langle
q|\slashed{J}_{1}|\frac{i\slashed{P}_{1}}{P_{1}^{2}}|\slashed{J}_{2}
|\cdots|\frac{i\slashed{P}_{k-1}}{P_{k-1}^{2}}|\slashed{J}_{k}|\bar{q}]~~~\Label{3.1-1}\end{equation}
where
$\slashed{J}_{i}=\gamma_{\mu}\cdot{J}_{i}^{\mu}\left(g_{i_{1}},...,g_{i_{k}}\right)$
is the contraction of gamma matrix and a off-shell gluon current
${J}_{i}^{\mu}\left(g_{i_{1}},...,g_{i_{k}}\right)$ and   the set
$\{1,2,...,n\}$ has been divided into sets $\{J_{i}\}$. After the
deformation, the  $z$-dependence is given as
\begin{equation} F\left(z\right)=\langle
q|\slashed{J}_{1}|\frac{i\slashed{P}_{1}+iz\slashed{l}}{P_{1}^{2}+2zP_{1}\cdot
l}|\slashed{J}_{2}|\cdots|\frac{i\slashed{P}_{k-1}+iz\slashed{l}}{P_{k-1}^{2}+2zP_{k-1}\cdot
l}|\slashed{J}_{k}|\bar{q}]~~~\Label{3.1-2}\end{equation}
where $\slashed{l}=q\rangle[\bar{q}$ is the null momentum used for
the deformation. Since $\lim_{z\rightarrow\infty} F(z)\sim z^0$, the
boundary contribution in (\ref{3.1-2}) is given by the value of
$F(z\to \infty)$, i.e.,
\bea F_{boundary}(z)= \langle
q|\slashed{J}_{1}|\frac{i\slashed{l}}{2P_{1}\cdot
l}|\slashed{J}_{2}|\cdots|\frac{i\slashed{l}}{2P_{k-1}\cdot
l}|\slashed{J}_{k}|\bar{q}] \eea
Summing up all possible contributions we finally get the boundary
term needed for the BCFW-recursion relation
\bea
A_{boundary} & = & \sum_{\{J_{i}\}}\langle q|\slashed{J}_{1}|
\frac{i\slashed{l}}{2P_{1}\cdot l}|\cdots|\frac{i\slashed{l}}
{2P_{k-1}\cdot l}|\slashed{J}_{k}|\bar{q}]\nonumber\\
 & = & \sum_{\{J_{i}\}}\langle q|\slashed{J}_{1}|\bar{q}]\cdot
 \prod_{1}^{k-1}\frac{i}{\langle q|P_{j}|\bar{q}]}\langle q|
 \slashed{J}_{j+1}|\bar{q}]~~~\Label{3-boundary-term}\eea
where the sum is over all possible splitting of $n$ gluons into $k$
sets with $k=1,...,n$ and the graphic representation is given in
Figure \ref{Fig:division}. Now we add the pole contribution and get
the full on-shell recursion relation with boundary contribution as
\bea
A\left(q^{-},\bar{q}^{+},g_{1},g_{2},...,g_{n}\right) & = &
\sum_{i=1}^{n-1}A\left(\WH{q}^{-},g_{1},...,g_{i},\WH{\bar{q}}_{\WH{P}}^{+}\right)
\cdot\frac{1}{P^{2}}\cdot A\left(\WH{q}_{-\WH{P}}^{-},g_{i+1},...,g_{n},\WH{\bar{q}}^{+}\right)\nonumber\\
 &  & +\sum_{\{J_i\}}\langle q|\slashed{J}_{1}|\bar{q}]\cdot\prod_{j=1}^{k-1}
 \frac{i}{\langle q|P_{j}|\bar{q}]}\langle q|\slashed{J}_{j+1}|\bar{q}]~~\Label{3-BCFW}\eea

The formula (\ref{3-BCFW}) is the main result of this subsection.
The pole part is given as sum of products of on-shell amplitudes
with lower points. The boundary part contains factors
$\Spab{q|\slashed{J}_{j+1}|\bar{q}}$\footnote{Pictorially  the
factor $\Spab{q|\slashed{J}_{j+1}|\bar{q}}$ represents the part of
amplitude $A(q^-, \{ J_{i}\}, \overline q^+)$ with pole $P_{J_{i}}$,
where to have momentum conservation, we need to redefine the
$\bket{q}$ and $\ket{\overline q}$.}, where the needed off-shell
current is discussed in previous subsection. It is worth to notice
that although each current is not gauge invariant, their sum gives
gauge invariant boundary contributions. Because this, sometimes a
good gauge choice could reduce the complexity of the calculation.
The gauge choice of each gluon must be consistent, i.e., same gauge
choice for all related current calculations. Here to exhibit the
details of the boundary recursion relation we give an explicit
example.

\subsection*{An example}
Using the recursion relation above, we calculate the 5-point QCD
amplitude and identify the results with that obtained directly by
Feynman diagrams. For a 5-point QCD amplitude with helicity
configuration
$A\left(1_{\bar{q}}^{+},2_{q}^{-},3^{-},4^{+},5^{+}\right)$, we
shift the momenta of $1_{q}^{-}$ and $2_{\bar{q}}^{+}$,\[
|1\rangle\rightarrow|1\rangle-z|2\rangle,\:|2]\rightarrow|2]+z|1]\]
and the recursion relation is given
\bea A\left(1_{\bar{q}}^{+},2_{q}^{-},3^{-},4^{+},5^{+}\right) & = &
A\left(\WH{2}_{q}^{-},3^{-},4^{+},\WH{P}_{\bar{q}}^{+}\right)\cdot\frac{1}{s_{23}}\cdot
A\left(-\WH{P}_{q}^{-},5^{+},\WH{1}_{\bar{q}}^{+}\right)\nonumber\\&&+
\sum_{\{J_i\}}\langle
q|\slashed{J}_{1}|\bar{q}]\cdot\prod_{j=1}^{k-1}\frac{i}{\langle
q|P_{j}|\bar{q}]}\langle q|\slashed{J}_{j+1}|\bar{q}]~. \eea
There is only one pole term, because a 4-point QCD amplitude
containing only one $-$ particle vanishes. And the whole boundary
contribution contains four terms $A^{b}_{\alpha}(\{J_i\})$ where
$\alpha=1,2,3,4$, with $\{J_i\}$ corresponding to
$\{J_{1}(3),J_{2}(4,5)\}$, $\{J_{1}(3,4),J_{2}(5)\}$,
$\{J_{1}(3,4,5)\}$ and $\{J_{1}(3),J_{2}(4),J_{3}(5)\}$.


We choose
the gauge as $q_{3}=k_{4},\: q_{4}=q_{5}=k_{3}$. Then the four terms
are given as
\bea
A^b_1\left(\{J_{1}(3),J_{2}(4,5)\}\right)&=&i\frac{[45]\langle32\rangle^{2}}
{\langle45\rangle\langle12\rangle s_{34}}\nonumber\\
A^b_2\left(\{J_{1}(3,4),J_{2}(5)\}\right)&=&i\frac{[12]\langle42\rangle\langle32\rangle}
{s_{34}\langle35\rangle\langle45\rangle}\nonumber\\
A^b_3\left(\{J_{1}(3,4,5)\}\right)&=&  0\nonumber\\
A^b_4\left(\{J_{1}(3),J_{2}(4),J_{3}(5)\}\right)&=&  i\frac{[13]^{2}\langle42\rangle^{2}}
{s_{34}\langle23\rangle\langle35\rangle[15]}\eea
Adding  the pole term
\[
A\left(\WH{2}_{q}^{-},3^{-},4^{+},\WH{P}_{\bar{q}}^{+}\right)
\cdot\frac{1}{s_{23}}\cdot A\left(-\WH{P}_{q}^{-},5^{+},\WH{1}_{\bar{q}}^{+}\right)
=i\frac{\langle23\rangle^{2}\langle35\rangle}{\langle34\rangle\langle45\rangle\langle25\rangle\langle15\rangle}
\]
we finally arrive
 \bea
A\left(1_{\bar{q}}^{+},2_{q}^{-},3^{-},4^{+},5^{+}\right)=i\frac{\langle23\rangle^{3}
\langle13\rangle}{\langle12\rangle\langle23\rangle\langle34\rangle\langle45\rangle\langle51\rangle}
\eea
which is same as  the result from Feynman diagram calculation.

From this example, we see that our calculation is a little bit
complicated than the one with gluon-pair deformation. However, the
point of this section is to provide a method to analyze boundary
contributions with fermion-pair deformation, which will be used in
next section.

\section{QCD amplitude with an anomalous magnetic moment}

Although BCFW recursion relation has been applied to many places,
for general effective field theories, the large $z$ behavior of the
amplitudes is not good enough to write down the original recursion
relation,  especially when the vertex contain momentum terms which
will  spoil the good large $z$ behavior. To deal with this problem,
there are several ideas one can try. One idea is to involve
auxiliary field to improve the large $z$ behavior
\cite{Benincasa:2007xk, Boels:2010mj}. The second idea is to replace
the problem by another equivalent theory with good behavior as did
in \cite{Larkoski:2010am}. The third idea is to use the boundary
BCFW recursion relation directly. In this part we will use the third
idea to study the effective theory of top quark with anomalous
magnetic moment couplings presented in \cite{Larkoski:2010am}.

Let us start with brief review of the theory with following
Lagrangian
 \bea
L=\bar{\Psi}\left[i\slashed{D}-m+\frac{ga}{4m}\Sigma_{\mu\nu}F^{\mu\nu
a}t^{a}\right]\Psi~~~\Label{Lag-Peskin}\eea
where $\Sigma_{\mu\nu}=\frac{i}{2}[\gamma_{\mu},\gamma_{\nu}]$ and
$a$ is the color index. As explained in \cite{Larkoski:2010am}, with
two gluon deformation $\Spba{g^+|g^-}$, $A(z)\to 0$ when $z\to
\infty$, thus the calculation of $q\bar{q}\to ng$ is reduced to the
case where all gluons are positive or negative helicities, which is
solved by an auxiliary scalar theory.

 The amplitude of this theory is a normal QCD
amplitude with several quark-gluon vertex replaced by the anomalous
magnetic moment vertex. For simplicity  we will set $m=0$, i.e., the
quark is  massless, and  focus on the case with $n$ gluons with
positive helicities $A(q,\bar{q};1^+,2^+,...,n^+)$. Since the
background field with only positive helicity gluons is self-dual,
the $\bar{\sigma}\cdot F$ piece of the magnetic momentum coupling is
zero and we are left with only ${\sigma}\cdot F$ piece. The nonzero
piece gives nonzero contribution when and only when fermion and
anti-fermion are all $+$ helicities. Considering the normal QCD
vertex is helicity-conserving, we conclude that for amplitude to be
nonzero, both external fermions must be $+$ helicity and there is
one and only one insertion of the magnetic moment coupling. After
the color ordering, the new vertex is just
\bea
V_{new}=\bar{\Psi}\cdot\frac{ga}{4M}\Sigma_{\mu\nu}F^{\mu\nu}\cdot\Psi~~~\Label{V-mag}\eea
which contains following   3-point vertex and a 4-point vertex
\bea
&&V_{new3}=\frac{ga}{2{\sqrt{2}}M}\cdot\left({p_{\mu}-\gamma_{\mu}\slashed{p}}\right)\nonumber\\
&&V_{new4}=\frac{ga}{8M}\cdot[\gamma_{\mu},\gamma_{\nu}]\eea

Now we discuss the large $z$-behavior of $A(q^+, 1^+,...,n^+,
\bar{q}^+)$ with  fermion momentum shifting
\bea
|q]\rightarrow|q]-z|\bar{q}],\;|\bar{q}\rangle\rightarrow|\bar{q}\rangle+z|q\rangle~~\Label{4.4}\eea
From (\ref{V-mag}) the new vertex won't infect the $z$ behavior
because momenta of gluons are not shifted. The fermion propagator
contributes $z^0$. For the fermion wave function, because of the new
vertex, $q$ and $\bar{q}$ now have the same helicities, thus unlike
the situation in previous section, we can't choose a proper
deformation to make the $z^0$-behavior from two wave-functions: the
best we can do is $z^1$-behavior.

Above $\lim_{z\to \infty} A(z)\to z^1$ behavior comes from naive
power counting, however, for the helicity configuration we are
interesting in, the result can be improved. To see it, we write down
a general expression from Feynman diagrams
 \bea
F=[q|\slashed{J}_{1}|\frac{i\slashed{P}_{1}}{P_{1}^{2}}|\slashed{J}_{2}|
\cdots|\frac{i\slashed{P}_i}{P^2_{i}}|\slashed{J}_i^{\star}|\cdots|
\frac{i\slashed{P}_{k-1}}{P_{k-1}^{2}}|\slashed{J}_{k}|\bar{q}]~~~\Label{P-F}\eea
where $\slashed{J}_{\alpha},\alpha\in\{1,...,i-1,i+1,...k\}$ stands
for the $\alpha$th current contracted with a normal QCD vertex, and
the current with a star $\slashed{J}_i^{\star}$ stands for the
current contracted with the anomalous magnetic moment vertex. Under
the deformation (\ref{4.4}), we have
\bea
F(z)=[q-zl|\slashed{J}_{1}|\frac{i\slashed{P}_{1}-iz\slashed{l}}
{P_{1}^{2}-2P_1\cdot zl}|\slashed{J}_{2}|\cdots|\frac{i\slashed{P}_i
-iz\slashed{l}}{P^2_{i}-2P\cdot zl}|\slashed{J}_i^{\star}|\cdots|
\frac{i\slashed{P}_{k-1}-iz\slashed{l}}{P_{k-1}^{2}-2P\cdot
zl}|\slashed{J}_{k}|\bar{q}]\Label{4.7} \eea
where the anomalous magnetic moment vertex contains both a 3-point
vertex and a 4-point vertex. As discussed in section two, a good
gauge choice with all $+$ helicities is that all  reference momenta
of external gluons are  same $r$. With this gauge choice, it's been
proven \cite{Larkoski:2010am} that the current
$J^{\mu}\left(1^+,2^+,...,m^+\right)$ contracted with magnetic
momentum coupling term could be written as
\bea
\slashed{J}_i^{\star}\left(1^+,2^+,...,m^+\right)=-i\frac{\left(1+\cdots+m\right)|r\rangle\langle
r|\left(1+\cdots+m\right)}{\langle
r1\rangle\langle12\rangle\cdots\langle mr\rangle}~~~\Label{P-J}\eea
and the current contracted with normal QCD 3-point vertex is given
by
 \bea
J_{\alpha}\left(i_{\alpha}^+,(i_{\alpha}+1)^+,...,({i_{\alpha+1}-1})^+\right)
&=&\frac{i\gamma_{\mu}}{\sqrt2}\cdot\frac{\langle r|\gamma^{\mu}
\slashed{P}_{i_{\alpha},(i_{\alpha+1}-1)}|r\rangle}{\sqrt{2}\langle
r{i_{\alpha}}\rangle\langle{i_{\alpha}}(i_{\alpha}+1)\rangle\cdots\langle
(i_{\alpha+1}-2),(i_{\alpha+1}-1)\rangle\langle (i_{\alpha+1}-1)r\rangle}\nonumber\\
&=&\sum_{t=i_{\alpha}}^{i_{\alpha+1}-1}\gamma_{\mu}\cdot\langle r|
\gamma^{\mu}|{p}_t]\langle{p}_tr\rangle A_{\alpha}
\eea
 Put back into (\ref{4.7}), we notice following typical combination
\bea [A|-iz\slashed{l}|\slashed{J}_{\alpha}|B]&=&iz[A|\bar{q}]
\langle q|\slashed{J}_{\alpha}|B]\nonumber\\
&=&iz[A|\bar{q}]\sum_{t=i_{\alpha}}^{i_{\alpha+1}-1}
\langle q|\gamma_{\mu}|B]\cdot\langle r|\gamma^{\mu}|{p}_t]\langle{p}_tr\rangle\nonumber\\
&=&iz[A|\bar{q}]2\langle
qr\rangle\sum_{t=i_{\alpha}}^{i_{\alpha+1}-1}[p_t|B]\langle
p_tr\rangle \eea
where Fietz identity has been used.  We find that if the gauge
choice is  $r=q$, this term will vanish. Similar thing will happen
when $-iz\slashed{l}$ is at the right side of
$\slashed{J}_{\alpha}$. By this analysis we see that power of $z$ in
numerator will be reduced. It is clear now that, for any diagram
that $k\geqslant3$ (i.e., the number of vertexes along the fermion
line) the large $z$-behavior  is good, while boundary contributions
with $z^0$ do appear  with $k=1,2$.

Based on above discussions,   the boundary BCFW recursion relation
is given as
 \bea
A^{\star}\left(q^{+},1^{+},2^{+},...,n^{+},\bar{q}^{+}\right)=\sum_{partition}A_{L}\cdot
\frac{1}{P^2}\cdot A_{R} +A_{boundary}~~\Label{P-BCFW}\eea
here $\star$ means that amplitude contains one anomalous magnetic
moment coupling. For our special helicity configuration, one of
$A_L, A_R$ will be the normal QCD amplitude, thus it could be
nonzero only for three-point amplitude and  the pole part is given
by %
 \bea \sum_{partition}A_{L}\cdot \frac{1}{P^2}\cdot A_{R}
=A\left(\WH{q}^+,1^+,\WH{\bar{Q}}_{P_{1q}}^-\right)\cdot \frac{1}{P_{1q}^2}
\cdot A^{\star}\left(\WH{Q}_{-P_{1q}}^+,2^+,...,\WH{\bar{q}}^+\right)\nonumber\\
+A^{\star}\left(\WH{q}^+,1^+,...,(n-1)^+,\WH{\bar{Q}}^+_{P_{n\bar{q}}}\right)
\cdot \frac{1}{P_{n\bar{q}}^2} \cdot A\left(\WH{Q}^-_{-P_{n\bar{q}}},n^+,\WH{\bar{q}}^+\right)
\eea
where $Q_P$ stands for a new involving quark with momentum $P$, and
$\star$ stands for an amplitude containing anomalous magnetic moment
coupling. In fact with the deformation (\ref{4.4}) only the second
term is nonzero.

Now we calculate the boundary contribution given by  two kinds of
Feynman diagrams with $k=1$ and $k=2$.
\begin{itemize}
\item $k=1$
\end{itemize}
There is only one Feynman diagram with $k=1$. It contains only one
vertex,  so it must be an anomalous magnetic moment. It's given as
\bea
A^b_1=[q|\slashed{J}^{\star}\left(1^+,2^+,...,n^+\right)
|\bar{q}]&=&-i\frac{[q|\slashed{P}_{1,n}|q\rangle\langle q|
\slashed{P}_{1,n}|\bar{q}]}{\langle q1\rangle\langle 12\rangle\cdots\langle nq\rangle}\nonumber\\
&=&-i\frac{[q\bar{q}]^2\langle q\bar{q}\rangle^2}{\langle q1\rangle\langle 12\rangle\cdots\langle nq\rangle}
\eea
where the momenta conservation $P_{1,n}+P_q+P_{\bar{q}}=0$ has been used.
\begin{itemize}
\item $k=2$
\end{itemize}
There are two kinds of Feynman diagrams of this type. In the  first
case, the anomalous vertex is connected next to the quark $q$. Their
contributions are given as
\bea
F_{\alpha}=\sum_{i=1}^{n-1}[q|\slashed{J}_1^{\star}\left(1^+,...,i^+\right)|
\frac{i\slashed{P}_{1,i}}{P^2_{1,i}}|\slashed{J}_2\left((i+1)^+,...,n^+\right)|\bar{q}]\eea
In the second case, the anomalous vertex is connected to the
antiquark $\bar{q}$. The contributions are given as
\bea
F_{\beta}=\sum_{i=1}^{n-1}[q|\slashed{J}_1\left(1^+,...,i^+\right)|
\frac{i\slashed{P}_{1,i}}{P^2_{1,i}}|\slashed{J}^{\star}_2\left((i+1)^+,...,n^+\right)|\bar{q}]\eea
 After the deformation, these two terms $F_{\alpha}(z)$, $F_{\beta}(z)$
  both behave as $\frac{z}{z}=z^0$. The boundary contributions can be
  calculated by the same way as in previous section and we get
\bea
A_2^b&=&\lim_{z\rightarrow+\infty}\left[F_{\alpha}(z)+F_{\beta}(z)\right]\nonumber\\
&=&\sum_{i=1}^{n-1}[q|\slashed{J}_1^{\star}
\left(1^+,...,i^+\right)|\bar{q}]\cdot\langle q|\slashed{J}_2\left((i+1)^+,...,n^+\right)|\bar{q}]\cdot\frac{1}{\langle q|P_{1,i}|\bar{q}]}\nonumber\\
&&+\sum_{i=1}^{n-1}[q|\slashed{J}_1
\left(1^+,...,i^+\right)|q\rangle \cdot [\bar{q}|\slashed{J}_2^{\star}\left((i+1)^+,...,n^+\right)|\bar{q}]\cdot\frac{1}{\langle q|P_{1,i}|\bar{q}]}\Label{A2}
\eea
Thus the whole boundary contribution is
\bea
A_{boundary}=A^b_1+A^b_2\eea

As we only consider about the case that all gluons have plus
helicity, there is another advantage we can take. As shown above,
the most convenient gauge choice for gluon currents with all plus
helicity is to choose all reference momenta to be a null vector $r$.
Go back to the recursion relation (\ref{P-BCFW}), both sides of the
equation should be gauge independent. Thus although $A^b_2$ contains
gauge dependent gluon currents, any gauge choice should give the
same result. So we can choose a special gauge which can simplify the
result. We find that if we choose $r=q$ \footnote{According to the
deformation $\Spab{\bar{q}|q}$, the gauge choice $r=\bar{q}$ won't
work at the same time}, then both terms in (\ref{A2}) vanish
\bea
\Spab{q|\slashed{J}_2\left((i+1)^+,...,n^+\right)|\bar{q}}&=&\frac{\Spab{q|\gamma_{\mu}|\bar{q}}}{\sqrt{2}}\cdot\frac{\Spaa{r|\gamma^{\mu}\slashed{P}_{i+1,n}|r}}{{\sqrt2}\Spaa{r,i+1}\Spaa{i+1,i+2}\cdots\Spaa{n-1,n}\Spaa{nr}}\nonumber\\
&=&\frac{\Spaa{qr}\Spab{r|\slashed{P}_{i+1,n}|\bar{q}}}{\Spaa{r,i+1}\Spaa{i+1,i+2}\cdots\Spaa{n-1,n}\Spaa{nr}}=0
\eea
and
\bea
[q|\slashed{J}_1
\left(1^+,...,i^+\right)|q\rangle&=&\frac{\Spab{q|\gamma_{\mu}|q}}{\sqrt2}\cdot\frac{\Spaa{r|\gamma^{\mu}\slashed{P}_{1,i}|r}}{{\sqrt2}\Spaa{r1}\Spaa{12}\cdots\Spaa{i-1,i}\Spaa{ir}}\nonumber\\
&=&\frac{\Spaa{qr}\Spab{r|\slashed{P}_{1,i}|\bar{q}}}{\Spaa{r1}\Spaa{12}\cdots\Spaa{i-1,i}\Spaa{ir}}=0
\eea
thus the second boundary term $A_2^b$ actually vanishes  with this
gauge choice and we are left with the final result:

\bea
A^{\star}\left(q^{+},1^{+},2^{+},...,n^{+},\bar{q}^{+}\right)=\sum_{partition}A_{L}\cdot
\frac{1}{P^2}\cdot A_{R} +A_{1}^b
\eea

\subsection*{An example}

Here we give an example with three positive helicity gluons using
the boundary BCFW recursion relation presented above. For such an
amplitude $A^{\star}\left(q^+,1^+,2^+,3^+,\bar{q}^+\right)$, the
recursion relation reads
\bea
A^{\star}\left(q^+,1^+,2^+,3^+,\bar{q}^+\right)&&=A(\WH{q}^+,1^+,\WH{\bar{P}}^-)\cdot\frac{1}{s_{1q}}\cdot A^{\star}(\WH{P}^+,2^+,3^+,\WH{\bar{q}}^+)\nonumber\\
&&+A^{\star}(\WH{q}^+,1^+,2^+,\WH{\bar{P}}^+)\cdot\frac{1}{s_{3\bar{q}}}\cdot A(\WH{P}^-,3^+,\WH{\bar{q}}^+)+A_1^b+A_2^b
\eea
where each term is given respectively
\bea
&&A(\WH{q}^+,1^+,\WH{\bar{P}}^-)\cdot\frac{1}{s_{1q}}\cdot A^{\star}
(\WH{P}^+,2^+,3^+,\WH{\bar{q}}^+)=-\frac{{\Spbb{\WH{q}1}}^3}
{\Spbb{1\WH{P}}\Spbb{\WH{P}\WH{q}}}\cdot\frac{1}
{s_{1q}}\cdot\frac{\Spbb{23}^2}{\Spaa{\WH{P}\WH{q}}}\nonumber\\
&&A^{\star}(\WH{q}^+,1^+,2^+,\WH{\bar{P}}^+)\cdot\frac{1}{s_{3\bar{q}}}\cdot A(\WH{P}^-,3^+,\WH{\bar{q}}^+)=-\frac{\Spbb{12}^2}{\Spaa{\WH{q}\WH{P}}}\cdot\frac{1}{s_{3\bar{q}}}\cdot\frac{\Spbb{3\WH{\bar{q}}}^3}{\Spbb{\WH{\bar{q}}\WH{P}}\Spbb{\WH{P}3}}\nonumber\\
&&A^b_1 = -\frac{\Spbb{q\bar{q}}^2\Spbb{q\bar{q}}^2}{\Spaa{q1}\Spaa{12}\Spaa{23}\Spaa{23}\Spaa{3\bar{q}}}\nonumber\\
&&A^b_2 = \frac{\Spbb{q1}\Spaa{qr}\Spab{r|1+q|\bar{q}}}{\Spaa{q1}\Spaa{r2}\Spaa{23}\Spaa{r3}} + \frac{\Spba{q|1+2|r}\Spab{r|1+2|\bar{q}}\Spaa{qr}}{\Spaa{r1}\Spaa{12}\Spaa{2r}\Spaa{r3}\Spaa{q3}}\nonumber\\
&&+\frac{\Spaa{qr}\Spbb{q1}\Spba{\bar{q}|2+3|r}\Spab{r|2+3|\bar{q}}}{\Spaa{r1}\Spaa{r2}\Spaa{23}\Spaa{3r}\Spaa{q1}\Spbb{1\bar{q}}} + \frac{\Spaa{qr}\Spab{r|3+\bar{q}|q}\Spbb{\bar{q}3}}{\Spaa{r1}\Spaa{12}\Spaa{2r}\Spaa{q3}}
\eea
Just as we've shown in the general case, the second boundary  term
$A_2^b$ vanishes if we set $r=q$. With this gauge choice, the
amplitude is simply given by
\bea
A^{\star}\left(q^+,1^+,2^+,3^+,\bar{q}^+\right)&=&-\frac{{\Spbb{\WH{q}1}}^3}
{\Spbb{1\WH{P}}\Spbb{\WH{P}\WH{q}}}\cdot\frac{1}
{s_{1q}}\cdot\frac{\Spbb{23}^2}{\Spaa{\WH{P}\WH{q}}}-\frac{\Spbb{12}^2}{\Spaa{\WH{q}\WH{P}}}\cdot\frac{1}{s_{3\bar{q}}}\cdot\frac{\Spbb{3\WH{\bar{q}}}^3}{\Spbb{\WH{\bar{q}}\WH{P}}\Spbb{\WH{P}3}}\nonumber\\
&&-\frac{\Spbb{q\bar{q}}^2\Spbb{q\bar{q}}^2}{\Spaa{q1}\Spaa{12}\Spaa{23}\Spaa{23}\Spaa{3\bar{q}}}\nonumber\\
&=&\frac{\Spbb{13}\Spbb{\bar{q}q}}{\Spaa{12}\Spaa{23}}+\frac{\Spbb{12}\Spbb{q1}}{\Spaa{23}\Spaa{\bar{q}3}}+\frac{\Spbb{23}\Spbb{\bar{q}3}}{\Spaa{q1}\Spaa{12}}
\eea
which numerically identifies to the result from naive Feynman diagram calculation.

 \section{Summary}\label{conclusion}

In this paper, we have presented  two main results. The first is the
BCFW recursion relation for off-shell gluon current. We show that we
can write down similar recursion relation with one modification: the
helicity sum of middle particle should over all four helicity states
instead of only two physical helicity states as familiar from our
BCFW recursion relation of on-shell amplitudes. For the off-shell
current, we have used two deformations. The first one is the
deformation with two on-shell external gluons. The second one is the
deformation with only one on-shell external gluon. For both
deformations, we must sum over all four helicity states to avoid the
gauge dependence of middle particle.

The second main result is how to  calculate boundary contributions
with deformed fermion pair by analyzing Feynman diagrams. We have
demonstrated our idea using two examples, the standard QCD and the
modified QCD. For modified QCD, we find that the actual large $z$
behavior under the deformation is better than naive power counting.
 Thus with the knowledge of off-shell gluon currents we give, the
boundary contributions can be calculated directly.

We must emphasize that our results in this paper is just a step
toward understanding the boundary contributions with general
deformations. There are  still a lot difficult questions waiting us
to investigate, for example, the property of zero raised in
\cite{Benincasa:2011kn}.

\subsection*{Acknowledgements}

 We are supported by fund from
Qiu-Shi, the Fundamental Research Funds for the Central Universities
with contract number 2010QNA3015, as well as Chinese NSF funding
under contract No.10875104, No.11031005.


\end{document}